\documentstyle[twoside,fleqn,espcrc2]{article}

\newcommand{\be}{\begin{equation}}
\newcommand{\ee}{\end{equation}}
\newcommand{\bea}{\begin{eqnarray}}
\newcommand{\eea}{\end{eqnarray}}
\newcommand{\bc}{\begin{center}}
\newcommand{\ec}{\end{center}}

\def\ut{\tilde{U}}
\def\mt{\tilde{M}}

\title{\bf A new way to deal with fermions in Monte Carlo simulations}

\author{T. Bakeyev
\address{Joint Institute for Nuclear Research, 141980 Dubna, Russia \\
email: bakeev@thsun1.jinr.ru}   
       }

\begin{document}

\begin{abstract}
An exact, nonlocal, finite step-size algorithm for Monte Carlo 
simulation of theories with dynamical fermions is proposed. 
The algorithm is based on obtaining the new configuration $\ut$
from the old one $U$ by solving the equation $ M(\ut) \eta= \omega M(U) \eta$,
where $M$ is fermionic operator,  
$\eta$ is random Gaussian vector, and $\omega$ is random real number 
close to unity. 
This algorithm can be used for acceleration of current  
simulations in theories with Grassmann variables. 
A first test was done for SU(3) QCD with purely fermionic term in the action. 
\vspace{1pc}
\end{abstract}

\maketitle


\section{Introduction.} 

From the earliest days of Monte Carlo simulations in lattice field theory
fermionic fields have caused annoying difficulty, which stems from their being 
anticommuting variables. It is not 
immediately straightforward to put fermions on a computer, which is expected 
to manipulate numbers. This problem is only an algorithmic one, since for 
the most actions in use one can eliminate fermions by an analytic integration. 
However, the resulting expressions involve the determinants
of very large matrices, making the numerical simulations extremely expensive.

Over the years many interesting tricks have been developed to circumvent this
problem. The most often used algorithm, Hybrid Monte Carlo (HMC) \cite{HMC}, 
is now considered to be the standart simulation tool. In recent years it 
was successfully challenged by the Multiboson method, which has many 
attractive features, like the possibility of simulation of an odd number 
of quark flavors. The most popular version of Multiboson method, 
proposed by M.~L\"uscher \cite{Luscher:1994xx}, has been intensively studied by
many authors (see \cite{Lutest} and references therein/thereon), and after
various improvements was claimed to be competitive or 
even slightly better than HMC
(see for example \cite{UV}). The Multiboson algorithm, proposed by 
A.~Slavnov \cite{Slavnov}, is less known, and was tested in \cite{SLtest}.
Besides HMC and Multiboson method, I like to mention
some interesting ideas, which may finally result in efficient algorithms 
for simulation of theories with fermions. They are based on 
the polymer \cite{polymer} and Jordan-Wigner occupation number \cite{OCCUP}
representations of fermion determinant,
the direct evaluation of Grassmann integrals \cite{CR2}, 
the separation of low and high eigenmodes of the Dirac operator \cite{Infrared},
and
the direct simulation of loop expansion by means of using the stochastic 
estimators \cite{NMC}.

Despite the variety of different approaches to the problem,
there is a common impression that all existing algorithms remain
inefficient \cite{Creutz}. This is clearly seen if we compare the 
computational cost for the theories with dynamical fermions from the one side, 
and purely bosonic
theories from the other side. In QCD even for heavy quarks the simulations  
are orders of magnitude slower than the ones in the quenched regime 
(the approximation
where the quark determinant is set to be equal to unity).  
Such a situation suggests that one should not stop the attempts to obtain 
relatively cheap fermion simulation algorithm.

In this paper I propose a new computational strategy for treating dynamical 
fermions in Monte Carlo simulations.
It has the virtues of exactness, nonlocality and finite step-size of update
\footnote{
There is a great probability, that a successful fermionic algorithm must 
be nonlocal 
and have a finite step-size of update. Indeed, HMC implies global changes
in configuration space, but with infinitesimal step-size. This drustically 
slows down
the simulations, becouse the high energy barries become nearly impassable.
Multiboson algorithm has the advantage of finite step-size, but it updates 
the fields locally. The large autocorrelation times 
suggest that such local changes are not very 'physical'
in this approach.}. 
Unlike the other exact algorithms,
the CPU-time required for the update in this algorithm grows
only lineary with the volume $V$ of the system.
This makes the new method to be rather attractive for 
practical simulations, although there are still some problems 
to be solved.

\section{The algorithm}

Suppose that we aim at sampling the partition function
of theory with two flavours of degenerate fermion fields: 
\be {\cal Z}_{ferm} = \int Det \;\Bigl( M^\dagger [U] M[U] \Bigr) \; dU \label{par_func}\ee
where $M$ is the discretized fermion operator acting at some vector space 
$\Omega$, $U$ denotes the bosonic degrees of freedom 
coupled to fermion fields. For simplicity of  
formulae I postpone the inclusion of purely bosonic action $S_b[U]$.
The sampling of full partition function  
\be {\cal Z}_{full} = \int e^{-S_b[U]} \; Det \;\Bigl( M^\dagger [U] M[U]
\Bigr) \; dU \label{full_par_f}\ee 
will be considered later.

At the beginning let me introduce a simple auxiliary algorithm,
which will guide the later construction and help to understand
the basis of new method. 
Starting from old configuration $U$, one executes the following 
instructions:

\bc {\it Prescription A} \ec
\begin{itemize}
\item generate random vector $\eta \in \Omega$ with Gaussian
distribution: 
$ \;\;\; P_G[\eta] = Z^{-1}_G e^{-|\eta|^2} $
\item propose new configuration $\ut$ with symmetric probability 
\be P_0[U \!\rightarrow \ut] = P_0[\ut \!\rightarrow U] \label{prop_0}\ee
\item accept $\ut$ with the probability 
\be P_{acc}[\eta ; U \!\rightarrow \ut] = 
min(1,e^{|\eta|^2 - |\mt^{-1}M\eta|^2}) \label{acc_p}\ee
\end{itemize} 
Here $\; Z_G$ normalizes the distribution: 
\be Z_G = \int e^{-|\eta|^2} d\eta d\eta^\dagger \ee
In eq.(\ref{acc_p}) and below I use the following short notations: 
$\mt \equiv M[\ut] ;\; M \equiv M[U]\;$. 

The full transition probability
\bea && P[U \!\rightarrow \! \ut] \equiv P_0[U \!\rightarrow \!\ut]\times
\nonumber \\ && 
\!\int\!\! d\eta d\eta^\dagger \; P_G[\eta]\; 
P_{acc}[\eta; \; U \!\!\rightarrow \!\ut\! ] 
\label{full_tr}\eea
satisfies the detailed balance condition 
with respect to the partition function (\ref{par_func}):
\be Det ( M^\dagger  M) P[U \rightarrow \ut] = 
Det ( \mt^\dagger  \mt) P[\ut \rightarrow U]\;\label{det_bal} \ee
This can be easily seen by making the change of variables
\be \eta \rightarrow M^{-1}\eta \quad ; 
\quad \eta^\dagger \rightarrow \eta^\dagger (M^{-1})^\dagger \label{change_m}\ee 
in the expression (\ref{full_tr}). Therefore, if the proposal matrix 
(\ref{prop_0}) ensures ergodicity, then {\it Prescription A} 
is a valid algorithm for sampling the partition function (\ref{par_func}).

However, this exact algorithm is expensive due to the necessity to 
invert the large matrix $\mt$ in the expression (\ref{acc_p}). The 
cost of inversion is proportional to the volume of the system $V$, so if
one updates the fields $U$ locally, the computational cost of 
updating the entire configuration grows as ${\cal O}(V^2)$. One 
can be more clever and perform the global updates by
fixing $\phi = M\eta$, introducing fictitious momenta $p$ conjugate
to the fields $\ut$ (or some functions of $\ut$), and pursuing a discrete 
integration of Hamilton evolution (this is the core of HMC algorithm,
which in its traditional form has a volume dependence like $V^{5/4}$). 
Nevertheless, the procedure remains expensive, becouse one should tend
the size of molecular dynamics steps to zero for minimizing the integration 
errors and again invert the matrix $\mt$ during the integration.  

Can one somehow modify the {\it Prescription A} to escape the inversion of 
matrix $\mt$, implementing the global update $U\rightarrow \ut$ 
in configuration space? The answer is 'yes'!

The key idea is to introduce into proposal matrix (\ref{prop_0})
the dependency on $\eta$ to make the new configuration $\ut$ satisfy 
the approximate equality 
$\mt^{-1}M\eta \approx \eta$, or equivalently,
\be \mt \eta \approx M \eta \label{approx_eq}\ee
The symmetry of expression (\ref{approx_eq}) under the interchange of 
variables $U\leftrightarrow \ut$ will help to garantee the 
reversibility of algorithm.
Moreover, the acceptance probability (\ref{acc_p}) should be large, if 
the approximate equation (\ref{approx_eq}) is close enough to its exact 
analogue.

After these handwaving speculations let me present a rigorous construction.
I do it in two steps. Firstly, I prove the detailed balance condition
for some general 
{\it Prescription B}, in which the proposal matrix 
$P_0[\omega,\eta\; ; \; U \!\rightarrow \ut]$ depends on $\eta$ and random
real number $\omega$ distributed in the narrow interval near unity.
Secondly, I specify the particular choice of 
$P_0[\omega,\eta\; ; \; U \!\rightarrow \ut]$ and present the final
algorithmic scheme.

\bc {\it Prescription B} \ec
\begin{itemize}
\item generate random vector $\eta \in \Omega$ with Gaussian
distribution: 
$ \;\;\; P_G[\eta] = Z^{-1}_G e^{-|\eta|^2} $
\item generate 
$\;\omega \in \Bigl[1-\epsilon\; ;\; \frac{1}{1-\epsilon}\Bigr]\;$
with the probability $\mu[\omega]$, satisfying 
\be  \frac{1}{\omega^2}\; \mu\Bigl[\frac{1}{\omega}\Bigr] =
\mu[\omega] \label{mu_omega}\ee
\item propose new configuration $\ut$ with the probability
$P_0[\omega,\eta\; ; \; U \!\rightarrow \ut]$, satisfying
the following symmetry relations: 
\be P_0[\omega,\eta\; ; \; U \!\rightarrow \ut] = 
P_0[{1\over \omega} ,\eta\; ; \; \ut \!\rightarrow U] 
\label{sym0}\ee
\bea && P_0[\omega,M^{-1}\eta\; ; \; U \!\rightarrow \ut] =
\nonumber \\ && =P_0[{1\over \omega},\mt^{-1}\eta\; ; \; \ut\!\rightarrow U]
\label{sym_db}\eea
\item accept $\ut$ with the probability (\ref{acc_p}).
\end{itemize} 
Here $\epsilon$ is the algorithmic parameter 
lying in the interval $0 \le \epsilon < 1$.

The condition (\ref{mu_omega}) provides the invariance of the measure
$\mu[\omega]d\omega$ under the change of variable 
$\omega \rightarrow {1\over \omega}$. Indeed, using eq.(\ref{mu_omega}), 
one can easily check that for 
any integrable function $f(\omega)$ the following equality holds:
\be 
\int_{1-\epsilon}^{\frac{1}{1-\epsilon}} \mu[\omega]\; f(\omega)\; d\omega
= \int_{1-\epsilon}^{\frac{1}{1-\epsilon}} 
\mu[\omega]\; f(1/\omega)\; d\omega  
\label{app_f}\ee

Using condition (\ref{sym0}) together with the property (\ref{app_f}),
one can demonstrate the symmetry of 
averaged over $\eta\; ,\omega$ proposal matrix
\bea && P_0^{av}[U\rightarrow \ut] \equiv 
\int_{1-\epsilon}^{\frac{1}{1-\epsilon}} \mu[\omega] d\omega 
\times \nonumber\\&& 
\int d\eta d\eta^\dagger \; P_G[\eta] \; 
P_0[\omega,\eta\; ; \; U \!\rightarrow \ut]
\label{p_0_av}\eea 
under the interchange of variables $U\leftrightarrow \ut$:
\be P_0^{av}[U\rightarrow \ut] = P_0^{av}[\ut \rightarrow U] 
\label{av_sym}\ee
This makes the algorithm reversible. 

Finally, condition (\ref{sym_db}) ensures 
the fulfillment of detailed balance equation (\ref{det_bal}) for 
the full transition probability 
\bea && P[U\rightarrow \ut] \equiv 
\int_{1-\epsilon}^{\frac{1}{1-\epsilon}} \mu[\omega] d\omega 
\int d\eta d\eta^\dagger P_G[\eta] \times \nonumber\\&&  \;
P_0[\omega,\eta\; ; \; U \!\rightarrow \ut]  \;
P_{acc}[\eta; \; U \!\!\rightarrow \!\ut\!]
\label{full_tr_p}\eea 
Indeed, making the change of variables
(\ref{change_m}) in expression (\ref{full_tr_p}), one obtains:
\bea && Det \; ( M^\dagger  M) \; P[U \rightarrow \ut] =
\int_{1-\epsilon}^{\frac{1}{1-\epsilon}} \mu[\omega] d\omega 
\times \nonumber\\&& 
Z^{-1}_G\int d\eta d\eta^\dagger \; min(e^{-|M^{-1}\eta|^2}, e^{- |\tilde{M}^{-1}\eta|^2}) 
\times \nonumber\\&& 
P_0[\omega,M^{-1}\eta\; ; \; U \!\rightarrow \ut]   \label{app_db}\eea
Expression (\ref{app_db})
is symmetric with respect to the interchange $U\leftrightarrow \ut$ 
due to the eqs.(\ref{sym_db},\ref{app_f}).
Therefore, {\it Prescription B} is again
a valid algorithm for sampling the partition function (\ref{par_func}),
if the averaged proposal matrix (\ref{p_0_av}) provides ergodicity.

Now I propose to specify the choice of the matrix 
$P_0[\omega,\eta\; ; \; U \!\rightarrow \ut]$ by defining it 
through the equation:
\be \mt \eta = \omega M\eta   
\label{basic}\ee
It means that, analytically or numerically, one finds some solution 
$\ut$ of the equation (\ref{basic}) and propose it as a new configuration.
A good recipe for the numerical search 
may be the local iterative minimization of 
the quantity 
\be R\equiv  |\; (\mt-\omega M)\eta \; |^2
\label{r_min}\ee  
for fixed $U,\omega,\eta$. Starting from $\ut = U$, 
the minimization proceeds until $R<\delta$ is reached, where $\delta$ 
determines the 
accuracy of solving the eq.(\ref{basic}).  
One should construct the minimization procedure in a way that ensures
reversibility of the algorithm (i.e. probability to obtain the 
configuration $\ut$ starting from $U$ at any $\omega$ should be equal
to probability to obtain $U$ starting from $\ut$ at $1/\omega$).

It can be checked that the proposal matrix, defined 
\footnote{The definition means the choice of some concrete
reversible procedure of finding the solution $\ut$.}
through the eq.(\ref{basic}),
satisfies the symmetry relation (\ref{sym0}). Indeed,
the equation (\ref{basic}) is invariant under the simultaneous interchange
of variables $U\leftrightarrow \ut; \; \omega \leftrightarrow 1/\omega$. 
Therefore, the proposals $P_0[\omega,\eta;  U \!\rightarrow \ut]$ and 
$P_0[1/\omega ,\eta ;  \ut \!\rightarrow U]$ are 
equiprobable, becouse they are defined through the same equation. 

The same logic is applicable for proving the fulfillment of symmetry 
relation (\ref{sym_db}). The lhs. of expression (\ref{sym_db}) is 
defined through the equation $\mt M^{-1}\eta = \omega\eta$,
meanwhile the rhs. is defined through the equation 
$M \mt^{-1}\eta = \frac{1}{\omega} \eta$. These equations are both equivalent
to the equation $M^{-1}\eta = \omega\mt^{-1}\eta$. Therefore,
the proposals $P_0[\omega,M^{-1}\eta ; \; U \!\rightarrow \ut]$ 
and $P_0[1/\omega,\mt^{-1}\eta ; \; \ut\!\rightarrow U]$
are also equiprobable. 

Let us note, that on the surface (\ref{basic}) the acceptance probability 
(\ref{acc_p}) acquires the following simple form:
\be P_{acc}[\omega,\eta] = 
min\; \bigl(1,e^{\bigl(1 - \frac{1}{\omega^2}\bigr)|\eta|^2}\bigr)
\label{acc_new}\ee 
The calculation of $P_{acc}$ becomes extremely cheap, since one does
not need to invert the matrix $\mt$ anymore. Moreover, the expression
(\ref{acc_new}) does not depend on $\ut$, so one can accept or reject
$\omega$ (or the pair $(\omega, \eta)\; $) even before solving 
the equation (\ref{basic}). 

We also need to specify the probability $\mu[\omega]$. 
A good choice is the following expression:
\be \mu[\omega] \propto
min ( 1,
1/\omega^2
) \label{prob_ch}\ee
which, evidently, satisfies the condition (\ref{mu_omega}).
Considering together the expressions (\ref{acc_new},\ref{prob_ch}),
one gets the unified probability distribution for $\omega$:
\bea && {\cal P}[\omega,\eta] =  \mu[\omega]\times P_{acc}[\omega,\eta]
\nonumber  \\ && \propto min\; \bigl(1/\omega^2, 
e^{\bigl(1 - \frac{1}{\omega^2}\bigr)|\eta|^2}\bigr) \eea
 
Now we are ready to write down the final algorithmic scheme 
for sampling the partition function (\ref{par_func}):

\bc {\it The algorithm} \ec
\begin{itemize}
\item generate random vector $\eta \in \Omega$ with Gaussian
distribution: 
$ \;\;\; P_G[\eta] = Z^{-1}_G e^{-|\eta|^2} $
\item generate 
$\;\omega \in \Bigl[1-\epsilon\; ;\; \frac{1}{1-\epsilon}\Bigr]\;$
with the probability: 
${\cal P}[\omega,\eta]\propto min\; \bigl(1/\omega^2, 
e^{\bigl(1 - \frac{1}{\omega^2}\bigr)|\eta|^2}\bigr)$ 
\item find the new configuration $\ut$ by solving the equation: 
$ \mt \eta = \omega M\eta $
\end{itemize} 

The only computationally expensive ingredient of the algorithm is 
the obtaining of solution $\ut$ of eq.(\ref{basic}). 
One can expect, that the usage of
the local iterative minimization of the functional (\ref{r_min})
gives the computational cost
proportional only lineary to the volume $V$.

If the procedure for obtaining the solution of eq.(\ref{basic})
is fully specified, the algorithm has only one free parameter
$\epsilon$, which controls the size of the deviation
of $\omega$ from unity. 
A rather intriguing possibility is to set $\epsilon = 0$,
therefore fixing $\omega = 1$. This makes the algorithm
'energy conserving', in a sense that $|\mt^{-1}M\eta|^2 = |\eta|^2$.
Unfortunately, in this case the search for the solution of eq.(\ref{basic})
by minimizing the functional (\ref{r_min}) does not work,
becouse one immediately obtains the trivial solution $\ut = U$. 
However, in typical case (e.g. SU(3) QCD), the solutions of eq.(\ref{basic})
are highly degenerate, and the subspace of nontrivial solutions in 
configuration space is not empty.
Finding the reversible procedure 
of nontrivial solving of equation $\mt\eta = M\eta$ is the perspective subject
for future investigations.

In gauge theories one can use the gauge freedom to simplify
the procedure of solving the eq.(\ref{basic}). Indeed, 
under the gauge transformations 
\be U\rightarrow U^g;\; \ut\rightarrow \ut^g\ee
eq.(\ref{basic}) acquires
the form: 
\be \mt \eta^g = \omega M\eta^g\label{basic_gt}\ee 
where $\eta^g\equiv G\eta$, $G$ is the matrix
representing the gauge transformation on the vector space $\Omega$. 
Solving the eq.(\ref{basic_gt})
can be particularly simple for some $\eta^g$. 

Finally let me note, that the same algorithm can be used 
for simulations in theories with bosonic determinants, if we 
change the acceptance (\ref{acc_new}). Sampling the 
partition function
\be {\cal Z}_{boson} = \int Det^{-1} \;\Bigl( M^\dagger [U] M[U] \Bigr) \; dU 
\label{par_bos}\ee 
one should use $ P_{acc}[\omega,\eta] = 
min\; (1,e^{\bigl(1 - \omega^2)|\eta|^2}\bigr)$
instead of the expression (\ref{acc_new}).

\section{Potential problems}

Despite the simple formulation and cheapness of the 
considered algorithm, it may be not applicable for some models. 
The main danger is the possible absence of solutions 
of eq.(\ref{basic}). This problem may be principal (no solutions
exist at all), or mild (no solutions exist for some particular
$\eta,\omega$). In the second case one can reject the pairs 
$(\eta, \omega)$ until the solution is found. 
The value of this 'hidden' acceptance would determine 
the actual efficiency of the algorithm in such a case.

Another possible problem may be connected with the procedure of finding
the solutions of eq.(\ref{basic}). Suppose that one uses the local iterative 
minimization of the quantity (\ref{r_min}) for this purpose.
If the functional (\ref{r_min}) can have some local minima
at which $R>0$, then minimization procedure can    
stick at some of these minima
before reaching $R=0$. In that case one should think of 
the other way for solving the eq.(\ref{basic}).

The ergodicity of the algorithm may also be under the question. One should 
check 
for the model of interest if any region of configuration space can be reached 
by the sequential updates via eq.(\ref{basic}). If the algorithm is nonergodic,
it can be used for the acceleration of other algorithms, like HMC and 
Multiboson. 

For given $U,\eta,\omega$ the eq.(\ref{basic}) can have many 
degenerate solutions 
\footnote{Such a situation takes place in SU(3) QCD, where the expression
(\ref{basic}) for unconditioned fermion matrix provides 
$(2*N_{dirac}*N_{color}*V) = 24V$ real equations for 
$(N_{generator}*4V) = 32V$ variables.}.
This is not a problem at all, if one 
respects the symmetry relations (\ref{sym0},\ref{sym_db})
when some particular solution is being chosen.
However, one should be careful in order not to violate the detailed
balance and reversibility.

In order to see, if these potential problems can cause any real troubles,
I made some tests for the case of SU(3) QCD with purely 
fermionic term \footnote{The choice of model is motivated by 
the popularity of computer simulations in QCD.}.
Let us however note, that these tests are preliminary, since for 
the reasons of simplicity and lack of computer time
I did not include pure bosonic part $S_b[U]$ into the action (this 
corresponds to $\beta = 0$). The more
complicated simulations for the full QCD in physically interesting region 
will be done in the future \cite{future}.

\section{Tests for SU(3) QCD with purely fermionic term in the action}

The simulations were performed at $8^4$ lattice for the partition function
(\ref{par_func}) with
\be M[U]=1 - k^2 D_{eo}D_{oe}\ee 
where $\; D[U]$ is usual Wilson difference operator
\bea && 
D_{xy} = \sum_{\mu} \; (1-\gamma_{\mu})\; U_{\mu}(x)\; \delta^{(4)}_{x+\hat{\mu} ,y}
+ \nonumber \\ && 
(1+\gamma_{\mu})\; U^{\dagger}_{\mu}(x-\mu)\;  \delta^{(4)}_{x-\hat{\mu} ,y}
\eea
and $D_{oe}$ means that this operator acts from even to odd space sites.
I used even-odd preconditioning to reduce the computational cost of the algorithm.
By doing so one also increases the degeneracy of solutions of 
eq.(\ref{basic}), which now provides $12V$ real equations for $32V$
unknown variables, since $M[U]$ acts only on even space manifold
\footnote{One may assume, that the more degenerate the solutions are,
the faster the configuration space is sampled, although this conjecture
should be confirmed in the future.}.

The hopping parameter $k$ was chosen to be $k=0.2$, which gives the plaquette 
value ${\cal P} = 0.0089(1)$. The performance of algorithm was compared
with that of usual HMC.
 
The minimization of functional (\ref{r_min}) was implemented by making 
the random moves in each color direction for all links lexicographically. 
Such minimization algorithm garantees reversibility of the procedure.

Firstly I checked the existence of solutions of eq.(\ref{basic}) at 
different $\omega$ for typical equilibrated configurations $U$.
The tests were done in the interval $\omega \in [0.9, 1.1]$. 
The good news is that for the
considered $\omega$ one always finds some solution 
(i.e. for
any $\delta$ the minimization procedure finally gives $R<\delta$). 
No local minima of the functional (\ref{r_min}) $\;\;R_{min} > 0$ 
were observed.

On the average solving the eq.(\ref{basic}) at $\epsilon = 0.05$ with precision 
$\delta = 5*10^{-7}$ required 42 minimization iterations 
(it was checked, that improving the precision did not affect the results).
Since the cost of one iteration is roughly equal to $2*N_{generator}=16$ 
multiplications 
by matrix $M[U]$, the total cost of finding the solution of
eq.(\ref{basic}) was approximately 670 matrix multilications. 
This has to be compared with the cost for generating one trajectory with HMC.
At step-size $\tau = 0.033$ and trajectory 
length equal to $1$ (this ensures 70\% acceptance),
one trajectory costed aproximately 
4420 matrix multiplications
\footnote{One trajectory included 33 conjugate gradient inversions 
of matrix $M[U]$. Each inversion costed approximately 134 
matrix multiplications.}, 
i.e. $\approx 6.6$ times more expensive than
solving eq.(\ref{basic}). 

Performing the tests it was observed that our algorithm
(let me name it Omega-algorithm in the following text) 
alone can not be used for
sampling the partition function (\ref{par_func}).
The problem is that probability 
${\cal P}[\omega,\eta]$ to accept $\omega<1$ is
strongly suppressed by large factor in the exponent. One knows that
the value of squared norm of Gaussian vector can be estimated as
$\; |\eta|^2 \sim I_{dof}$, where $ I_{dof} $ is the dimensionality
of vector space on which $\eta$ is defined. 
For even-odd preconditioned SU(3) QCD at $8^4$ lattice one has 
$I_{dof} = 24576$! 
Therefore, using Omega-algorithm alone,
one is restricted to choose between two unfavorable possibilities: \\
1) Using $\epsilon \sim 1/I_{dof} $. 
It was observed that at such small values of $\epsilon$
the evolution of $U$ fields becomes 
very slow, becouse the new configuration $\ut$ always lies too close to
the old one (large autocorrelation times);\\
2) Using $\epsilon \gg 1/I_{dof}$. Then $\omega < 1$ is almost never 
accepted, and the system is gradually cooled until it does not move.

In the second case the trouble appears due to the nonergodicity of algorithm 
at large $\epsilon$. Nevertheless, even at large $\epsilon$
Omega-algorithm can be used for acceleration of other fermionic 
algorithms like HMC or Multiboson, since the ergodicity is provided 
by them.

I tested the combination of Omega-algorithm and HMC. After each
HMC trajectory the global move in gauge configuration space was
implemented
by using Omega-algorithm at $\epsilon = 0.05$. The average plaquette value
for this algorithmical mixture was ${\cal P} = 0.0091(1)$, 
coinciding with correct one within the error bars. The efficiency of 
HMC-Omega mixture was estimated by measuring
the autocorrelation times for the plaquette. The results were: \\
Pure HMC, 1000 trajectories: $\tau_{int}=1.4(2)$; \\
HMC+Omega, 1000 (traj + $\omega$-update): $\tau_{int}=0.7(1)$. \\
Note, that autocorrelation time was reduced almost at no cost, 
becouse the global Omega-algorithm update is much cheaper, than HMC
trajectory.  
Of course, these tests are not very illustrative, becouse the 
autocorrelation times for theory with pure fermionic term at this kappa
value are rather small. 
The further tests in models of practical relevance are needed.
In particular, one should 
find a cheap way of inclusion of pure bosonic action $S_b[U]$ in simulations.
This issue is discussed in the next section.

Let me summurize the results of the first test for Omega-algorithm. In its present
version the algorithm was found to be nonergodic. Nevertheless, it was
demonstrated, that Omega-algorithm can be efficiently used for acceleration 
of other fermionic algorithms. 

Here I want to emphasize, that nonergodicity is not intrinsic 
for Omega-algorithm.
It can be rather attributed to the unfavorable procedure of finding the
solution of eq.(\ref{basic}) by minimizing the functional (\ref{r_min}), 
which forces us to use large $\epsilon$.
One can still hope that it is possible to use 
$\epsilon \sim1/I_{dof}$, or even $\epsilon = 0$, 
and obtain nontrivial solutions $\ut$ far away from $U$ at the same time. This hope
is based on the high degeneracy of solutions of eq.(\ref{basic}). The main problem
is to garantee the reversibility of the procedure. An investigation along these lines
is in progress \cite{future}.

\section{Inclusion of bosonic action} 

Now let us consider the sampling of full partition function 
(\ref{full_par_f}). 
The simplest way to include the contribution of purely bosonic sector
is to add to the algorithm of the previous section 
the following instruction:

\begin{itemize}
\item accept the new configuration $\ut$ with the probability
\be P^{boson}_{acc} = min\; \bigl(1, e^{S_b[U]-S_b[\ut]}\bigr) 
\label{boson_acc}\ee
\end{itemize} 
The reader can easily check that one gets a valid algorithm for 
sampling the partition function (\ref{full_par_f}). 
However, the new configuration $\ut$, obtained by solving the 
eq.(\ref{basic}), differs globally from the old one,
so the acceptance (\ref{boson_acc}) can be very small. Of course,
one can try to make the new
configuration lying close to the old one to ensure the reasonable 
acceptance, but by doing so one lose one of the main 
advantages of the new algorithm - the finite step-size of update. 

A more radical way is to rewrite
the partition function (\ref{full_par_f}) in the form:
\be {\cal Z}_{full}  = \int Det \;\Bigl( B^\dagger [U] B[U] \Bigr) \; dU 
\label{mod_par_f}\ee
where operator $B$ satisfies the following identity:
\be Det\; B[U] = e^{-S_b[U]/2} Det\; M[U]\ee 
After that one can use the  algorithm of the previous section
with the equation 
\be B[\ut] \eta = \omega \; B[U]\eta \label{basic_2}\ee
instead of eq.(\ref{basic}). There is a large degree of 
ambiguity in defining the operator $B$. For example,
one can choose $B[U] = e^{-S_b[U]/2N} M[U]$, where $N$ is the 
size of matrix $M$. A more rational way is, probably, to insert  
the bosonic contribution into the operator $M$ locally.
One can represent the bosonic action as a sum over local contributions:
\be \frac{1}{2} S_b[U] = \sum_{i=1}^{N} S_b^{(i)}[U] \ee
(some of $S_b^{(i)}$ can be equal to zero), 
and then define the matrix B as follows:
\be B_{ij} =  e^{-S_b^{(i)}} M_{ij}\; 
\qquad 1\leq i,j \leq N\label{b_op}\ee   
(no summation over $i$ is assumed in (\ref{b_op})). In any case, 
one should try to choose the most convenient version of operator
$B$ in order to simplify the procedure of solving the eq.(\ref{basic_2}).

\section{Discussion.}

The new method, proposed in this paper,
can provide a cheap simulation algorithm for theories 
with dynamical fermions. The algorithm is exact 
and has a finite step-size of update.

Although in its present version the algorithm was found to be nonergodic, 
it can be used in combination with other ergodic algorithms like HMC and
Multiboson. One may expect a good performance for such algorithmical mixture
when the lattice volume increases, since computational cost for our
new algorithm 
grows only lineary with the volume of system. 

The efficiency of algorithm can be strongly affected
by the clever choice of procedure of solving the eq.(\ref{basic}).
Due to the high degeneracy of solutions 
one may hope to find the procedure, which makes the algorithm ergodic,
and samples the configuration space fast enough.
Also one can speculate on possibility of finding some analytic
solutions for the eq.(\ref{basic}). If this is feasible
at all, one can obtain a very cheap fermionic algorithm,
comparable in cost with the algorithms for purely bosonic theories.

\vspace{0.3cm}

{\bf Acknowledgments:}
I am grateful to Ph. de Forcrand for helpful comments. 
I also like to thank A. Slavnov and K. Jansen for interesting
discussions.
This work was supported by Russian Basic Research Fund 
under grant \# 99-01-00190 and grant for young scientists
\# 01-01-06089.

\end{document}